\documentclass[longauth]{aa}

\usepackage{graphicx}
\usepackage{txfonts}
\PassOptionsToPackage{hyphens}{url}\usepackage[breaklinks]{hyperref}

\newcommand{\orcit}[1]{\protect\href{https://orcid.org/#1}{\protect\includegraphics[width=8pt]{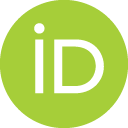}}}

\newcommand{\nbTargets}{134}
\newcommand{\nbTargetsSophie}{72}
\newcommand{\nbTargetsUves}{70}

\newcommand\gaia{\textit{Gaia}}

\newcommand\gdrtwo{\gaia~DR2}
\newcommand\gedrthree{\gaia~EDR3}
\newcommand\gdrthree{\gaia~DR3}
\newcommand\gdrfour{\gaia~DR4}
\newcommand\gdrfive{\gaia~DR5}

\newcommand{\Teff}{T_{\rm eff}}
\newcommand{\logg}{\log g}

\newcommand\ms{\ensuremath{\text{m~s}^{-1}}}
\newcommand\kms{\ensuremath{\text{km~s}^{-1}}}

\newcommand{\mygi}{MyGIsFOS}

\usepackage{xcolor}
\nolinenumbers
\begin{document}

   \title{\gdrthree\ high radial velocity stars: Genuine fast-moving objects or outliers?
   \thanks{Based on observations made with SOPHIE at OHP and UVES at VLT.}}

   \author{D.~Katz\orcit{0000-0001-7986-3164}\inst{\ref{lira}}
          \and A.~G\'omez\inst{\ref{lira}}
          \and E.~Caffau\orcit{0000-0001-6011-6134}\inst{\ref{lira}}
          \and P.~Bonifacio\orcit{0000-0002-1014-0635}\inst{\ref{lira}}
          \and C.~Hottier\orcit{0000-0002-3498-3944}\inst{\ref{unidia}}
          \and O.~Vanel\inst{\ref{unidia}}
          \and C.~Soubiran\orcit{0000-0003-3304-8134}\inst{\ref{bordeaux}}
          \and P.~Panuzzo\orcit{0000-0002-0016-8271}\inst{\ref{unidia}}
          \and D.~Chosson\orcit{0009-0009-2092-5514}\inst{\ref{unidia}}
          \and P.~Sartoretti\inst{\ref{unidia}}
          \and R.~Lallement\inst{\ref{lira}}
          \and P.~Di~Matteo\orcit{0000-0002-5213-4807}\inst{\ref{lira}}
          \and M.~Haywood\orcit{0000-0003-0434-0400}\inst{\ref{lira}}
          \and N.~Robichon\orcit{0000-0003-4545-7517}\inst{\ref{lira}}
          \and S.~Baker\orcit{0000-0002-6436-1257}\inst{\ref{mssl}}
          \and A.~Barbier\orcit{0009-0004-0983-931X}\inst{\ref{cnes}}
          \and D.~Bashi\orcit{0000-0002-9035-2645}\inst{\ref{cambridge}}
          \and K.~Benson\inst{\ref{mssl}}
          \and R.~Blomme\orcit{0000-0002-2526-346X}\inst{\ref{brussels}}
          \and N.~Brouillet\orcit{0000-0002-3274-7024}\inst{\ref{bordeaux}}
          \and L.~Casamiquella\orcit{0000-0001-5238-8674}\inst{\ref{lira}}
          \and L.~Chemin\orcit{0000-0002-3834-7937}\inst{\ref{strasbourg}}
          \and M.~Cropper\orcit{0000-0003-4571-9468}\inst{\ref{mssl}}
          \and Y.~Damerdji\orcit{0000-0002-3107-4024}\inst{\ref{alger},\ref{liege}}
          \and C.~Dolding\orcit{0009-0003-7199-6108}\inst{\ref{mssl}}
          \and S.~Faigler\inst{\ref{telaviv}}
          \and Y.~Fr\'{e}mat\orcit{0000-0002-4645-6017}\inst{\ref{brussels}}
          \and E.~Gosset\inst{\ref{liege},\ref{fnrs}}
          \and A.~Guerrier\inst{\ref{cnes}}
          \and R.~Haigron\inst{\ref{unidia}}
          \and H.E.~Huckle\inst{\ref{mssl}}
          \and N.~Leclerc\orcit{0009-0001-5569-6098}\inst{\ref{unidia}}
          \and A.~Lobel\orcit{0000-0001-5030-019X}\inst{\ref{brussels}}
          \and O.~Marchal\orcit{0000-0001-7461-8928}\inst{\ref{strasbourg}}
          \and T.~Mazeh\inst{\ref{telaviv}}
          \and A.~Mints\orcit{0000-0002-8440-1455}\inst{\ref{aip}}
          \and F.~Royer\orcit{0000-0002-9374-8645}\inst{\ref{lira}}
          \and G.~M.~Seabroke\orcit{0000-0003-4072-9536}\inst{\ref{mssl}}
          \and M.~Smith\inst{\ref{mssl}}
          \and O.~Snaith\orcit{0000-0003-1414-1296}\inst{\ref{exeter}}
          \and F.~Th\'{e}venin\inst{\ref{oca}}
          \and K.~Weingrill\orcit{0000-0002-8163-2493}\inst{\ref{aip}}
          }

   \institute{LIRA, Observatoire de Paris, Universit\'e PSL, Sorbonne Universit\'e, Universit\'e Paris Cit\'e, CY Cergy Paris Universit\'e, CNRS, 92190 Meudon, France\relax\label{lira}
   \and
   UNIDIA, Observatoire de Paris, Universit\'e PSL, CNRS, 92190 Meudon, France\relax\label{unidia}
   \and
   Laboratoire d'astrophysique de Bordeaux, Universit\'{e} de Bordeaux, CNRS, B18N, all{\'e}e Geoffroy Saint-Hilaire, 33615 Pessac, France\relax\label{bordeaux}
   \and
   Mullard Space Science Laboratory, University College London, Holmbury St Mary, Dorking, Surrey RH5 6NT, United Kingdom\relax\label{mssl}
   \and
   CNES Centre Spatial de Toulouse, 18 avenue Edouard Belin, 31401 Toulouse Cedex 9, France\relax\label{cnes}
   \and
   Astrophysics Group, Cavendish Laboratory, University of Cambridge, JJ Thomson Avenue, Cambridge CB3 0HE, UK\relax\label{cambridge}
   \and
   Royal Observatory of Belgium, Ringlaan 3, 1180 Brussels, Belgium\relax\label{brussels}
   \and
   Universit\'{e} de Strasbourg, CNRS, Observatoire Astronomique de Strasbourg, UMR 7550, 67000 Strasbourg, France\relax\label{strasbourg}
   \and
   CRAAG - Centre de Recherche en Astronomie, Astrophysique et G\'{e}ophysique, Route de l'Observatoire Bp 63 Bouzareah 16340, Alger, Alg\'erie\relax\label{alger}
   \and
   Institut d'Astrophysique et de G\'{e}ophysique, Universit\'{e} de Li\`{e}ge, 19c, All\'{e}e du 6 Ao\^{u}t, B-4000 Li\`{e}ge, Belgium\relax\label{liege}
   \and
   School of Physics and Astronomy, Tel Aviv University, Tel Aviv 6997801, Israel\relax\label{telaviv}
   \and
   F.R.S.-FNRS, Rue d'Egmont 5, 1000 Brussels, Belgium\relax\label{fnrs}
   \and
   Leibniz Institute for Astrophysics Potsdam (AIP), An der Sternwarte 16, 14482 Potsdam, Germany\relax\label{aip}
   \and
   University of Exeter, School of Physics and Astronomy, Stocker Road, Exeter, EX4 4QL, UK\relax\label{exeter}
   \and
   Universit\'{e} C\^{o}te d'Azur, Observatoire de la C\^{o}te d'Azur, CNRS, Lagrange UMR 7293, CS 34229, 06304, Nice Cedex 4, France\relax\label{oca}
             }

   \date{Received TBD; accepted TBD}
 
  \abstract
   {The third \gaia\ data release includes 33.8 million radial velocity measurements, extending to a magnitude of $G_\mathrm{RVS} = 14$. To reach this magnitude limit, spectra were processed down to a signal-to-noise ratio (S/N) of 2. In this very low S/N regime, noise-induced peaks in the cross-correlation function can result in spurious radial velocity determinations. Quality filters were applied to the dataset to mitigate such artefacts as much as possible prior to publication. Nevertheless, the high radial velocity (HRV) stars -- defined here as those with radial velocities below $-$500 or above $+$500 \kms -- are so sparsely populated that even a few hundred spurious measurements can lead to significant contamination.}
   {The objectives of the present study are as follows: (i) to confirm or refute the radial velocity values of the order of one hundred \gdrthree\ HRV stars, (ii) to evaluate the rate of spurious radial velocities in the \gdrthree\ catalogue as a function of S/N and radial velocity, and (iii) to examine the properties of the genuine HRV stars.}
   {A total of 134 \gdrthree\ HRV stars were observed using the SOPHIE and UVES spectrographs. Their radial velocities were determined via cross-correlation or template-matching methods. These measurements were subsequently combined with radial velocities from the APOGEE, GALAH, GES, LAMOST, and RAVE catalogues in order to assess the rate of erroneous \gdrthree\ radial velocities as a function of S/N and radial velocity range. Finally, the orbits of a clean sample of HRV stars were integrated using an axisymmetric Galactic gravitational potential.}
   {Ground-based measurements confirm the \gdrthree\ radial velocities of 104 out of our 134 targets, and they refute those of the remaining 30. The combination of these data with the spectroscopic surveys mentioned above enabled an assessment of the rate of spurious measurements as a function of S/N and across three intervals of absolute value of the radial velocity: [0, 200), [200, 400), and [400, 1000)~\kms. The outlier rate reaches up to 83\% in the S/N range [2, 3) and velocity interval [400, 1000)~\kms, and it decreases rapidly with increasing S/N and/or with decreasing absolute value of the radial velocity. The confirmed radial velocities were then combined with \gdrthree\ HRV stars having S/N~>~10, in order to construct a clean sample of HRV stars. The majority of these stars follow retrograde orbits. Their location in the energy-vertical component of the angular momentum diagram coincides with the region where several structures associated with past merging events have been identified: Sequoia, Arjuna and I'itoi, Antaeus, ED-2, and ED-3. It is likely that most of these HRV stars were accreted.}
   {}

   \keywords{Catalogues - Techniques: radial velocities - Stars: kinematics and dynamics}

   \maketitle

\nolinenumbers

\section{Introduction}
The interest for fast-moving stars is approximately a century old (e.g. Oort, 1926); although, the threshold above which an object is considered fast has radically changed over time. Nowadays, the quest for high- or hyper-velocity stars (HVSs) is more active than ever \citep{Brown2005, Brown2014, Brown2018, Boubert2018, Marchetti2019, Marchetti2022, Koposov2020, Quispe-Huaynasi2022, Quispe-Huaynasi2023, Quispe-Huaynasi2024, Li2023, Liao2023, Scholz2024, Verberne2024, Carballo-Bello2025}, and supported by large public surveys, such as \gaia\ \citep{GaiaCollaboration2016, GaiaCollaboration2018, GaiaCollaborationBrown2021, GaiaCollaborationVallenari2023}, LAMOST \citep{Cui2012, Zhao2012}, and APOGEE \citep{Majewski2017, Abdurrouf2022}, in particular. HVSs are of interest in several astrophysical domains. They can probe the extreme dynamical processes that can accelerate stars \citep[see][for a review]{Brown2015}, and in particular the interaction with the Milky Way central massive black hole, Sagittarius A* \citep[e.g.][]{Hills1988, Brown2005, Koposov2020, Evans2022, Evans2023, Liao2023, Verberne2024, Hattori2025}, and the Large Magellanic Cloud central black hole \citep[e.g.][]{Evans2021, Han2025}. They are also used to constrain the Milky Way escape velocity \citep[e.g.][]{Monari2018}, to characterise its dark matter halo \citep[e.g.][]{Contigiani2019, Gallo2022} and to measure its mass \citep[e.g.][]{Monari2018}, and they constitute key fossil records of the early phase of the Milky Way assembly as well as of the halo accretion history \citep[e.g.][]{Caffau2020, Huang2021, Li2022, Bonifacio2024}.

Finding HVSs is a difficult task, because they are very rare objects distributed in the very large volume of the Milky Way bulge, disc, and halo. The third \gaia\ data release (\gdrthree), published on 13~June~2022 \citep{GaiaCollaborationVallenari2023}, contains, among others, 33.8 million radial velocities \citep{Katz2023, Blomme2023} down to a $G_\mathrm{RVS}$ magnitude of 14. Most of these stars also have published positions, parallaxes, and proper motions in \gedrthree\ \citep{GaiaCollaborationBrown2021, Lindegren2021astrometry, Lindegren2021bias}. The raw number and full sky coverage of the combined astro-spectroscopic sample represent two very strong assets to find new HVSs. However, to reach magnitude 14, radial velocities have been derived down to signal-to-noise ratios (S/Ns) as low as S/N = 2 per pixel. In this very low S/N regime, false secondary cross-correlation peaks (CC peaks) can sometimes exceed the true primary CC peak and produce a spurious radial velocity. Other issues, such as nearby bright stars \citep{Boubert2019, Seabroke2021}, a template mismatch or a processing problem, can also spoil the measurements. Many quality filters have been applied to the data and the majority of the spurious radial velocities have been discarded \citep{Katz2023}. The remaining erroneous values represent a small fraction of the full dataset. Yet, the radial velocity tails (typically below or above $\pm$400 or 500 \kms) are so sparsely populated that even a few hundred spurious measures can significantly contaminate them. As a consequence, in \gdrthree, the stars with high radial velocities should be considered with caution. This is of course an issue, as the false large radial velocities erroneously mimic HVSs.

The first objective of the present study is to observe, with ground-based spectrographs, of the order of a hundred stars with high \gdrthree\ radial velocities (absolute values larger than 500 \kms), in order to confirm or refute the \gaia\ measurements. The selection of the targets is presented in Sect.~\ref{Sect:Data}. Sect.~\ref{Sect:Measure} describes the methods used to measure the radial velocities. In Sect.~\ref{Sect:Confirmed}, we explain how ground-based velocities were used to confirm or refute the \gaia\ values. The confirmed and refuted groups were subsequently combined with public ground-based spectroscopic surveys to assess the rate of spurious radial velocities in \gdrthree\ as a function of S/N (Sect.~\ref{Sect:OutliersRate}). These statistics were subsequently used to select a clean sample of high radial velocity stars. Interestingly, it turns out that most of it consists of stars with retrograde galactocentric azimuthal velocities. This is discussed in Sect.~\ref{Sect:Properties}. We conclude in Sect.~\ref{Sect:Conclusion}.

\section{Data\label{Sect:Data}}
\subsection{Target selection, observation, and reduction\label{Sect:Targets}}
The targets were selected from the 1544 \gdrthree\ stars with an absolute value of the radial velocity larger than 500 \kms\ (hereafter referred to as HRV). They were chosen to be sufficiently bright for the instruments employed, to span a large range of \gdrthree\ S/N, from the lower limit of 2 up to above 100, and to ensure a balanced distribution of stars with both negative and positive \gdrthree\ radial velocities. In total \nbTargets\ stars were observed: \nbTargetsSophie\ stars with the northern hemisphere SOPHIE spectrograph \citep{Perruchot2008} mounted on the 193~cm telescope of the Observatoire de Haute-Provence and \nbTargetsUves\ stars with the southern hemisphere UVES spectrograph \citep{Dekker2000} at VLT. Eight stars were observed both with SOPHIE and UVES. The SOPHIE spectra were collected in visitor mode during two runs: 5 nights from 27 to 31 July 2022 and 4 nights from 16 to 19 January 2023. SOPHIE is a fibre-fed echelle spectrograph covering the wavelength range 387--694~nm split over 39 orders. To maximise the signal-to-noise ratio, the spectrograph was operated in high-efficiency mode (resolving power 39~000) and slow CCD readout mode. The exposure time was adapted to the magnitude of the targets, from about 5 minutes for the brightest ($V < 12 $~mag) up to 1 hour for the faintest ($V \in [14.25, 14.75]$~mag). The spectra were reduced during the observing runs with the standard SOPHIE reduction pipeline. The UVES spectra were collected in service mode during the ESO periods P109, P110 (1 April 2022--31 March 2023) and P112 (1 October 2023--31 March 2024). The spectrograph was operated with the setting DIC2 437+760 (wavelength ranges 373–-499 and 565–-946 nm), with a 0\farcs{4} slit (resolving power 90~000) in the blue arm and 0\farcs{3} (resolving power 110~000) in the red arm. The spectra were reduced with the ESO pipeline.

The targets are presented in Table~A.1, along with their \gdrthree\ right ascensions, declinations, and $G$ magnitudes. For each observation, the table additionally lists the spectrograph employed and the Modified Julian Date (MJD) corresponding to the start of the exposure. Throughout this paper, we refer to the stars using  short names, which are more readable than the longer \gaia\ \verb|source_id|. We use \verb|Gaia_| followed by one to four digits for SOPHIE observations, and \verb|RVS| followed by three or four digits for UVES observations. The mapping between these short names and the corresponding \gaia\ \verb|source_id| is also provided in Table~A.1. Table~A.1 is available online and is described in Appendix~\ref{App:Targets}.

\subsection{Metallicities of the stars observed with SOPHIE}
The atmospheric parameters and chemical compositions of the stars observed with the UVES spectrograph have already been published and discussed in two companion papers: \citet{Caffau2024} and \citet{Caffau2025}. In this paper we provide atmospheric parameters and metallicities derived for 50 stars observed with SOPHIE, with a high-enough S/N. Effective temperatures and surface gravities were derived from the Gaia photometry and parallaxes as described in \citet{Caffau2024}. The photometry was corrected for reddening using the 3D extinction maps of \citet{Vergely2022}. The parallaxes were corrected for the zero-point as described in \citet{Lindegren2021bias}. The metallicities were derived using \mygi\ \citep{Sbordone2014} on the SOPHIE spectra using the grids of synthetic spectra  described in \citet{Caffau2021}. The atmospheric parameters and metallicities for these 50 stars are provided in Table~B.1, which is available online and is described in Appendix~\ref{App:AP}.

\subsection{Ground-based surveys\label{Sect:Surveys}}
To complement the sample of \nbTargets\ HRV candidates observed with SOPHIE and UVES, we cross-matched the 1544 \gdrthree\ HRV, with five ground-based spectroscopic surveys, namely: APOGEE~DR17 \citep{Abdurrouf2022}, GALAH~DR3 \citep{Buder2021, Zwitter2021}, the GAIA-ESO Survey (GES) DR3 \citep[see for the survey description,][]{Gilmore2012, Randich2013}, LAMOST~DR7 \citep[see for the survey and pipeline description,][]{Zhao2012, Deng2012, Luo2015}, and RAVE~DR6 \citep{Steinmetz2020_1, Steinmetz2020_2}. After applying quality cuts on the ground-based surveys (see Appendix~\ref{App:GBS}), we found 49 matches: 23 in APOGEE, 7 in GALAH, 0 in GES, 15 in LAMOST and 4 in RAVE. Two of them, one APOGEE star and one LAMOST star, were also observed with UVES and SOPHIE respectively and are among the \nbTargets\ HRV candidates. In Figs.~\ref{Fig:Data}, \ref{Fig:Outliers1} and \ref{Fig:Properties1}, GBS refers to the 49 ground-based survey stars minus our two HRV candidates.

\subsection{Radial velocities versus Galactic longitudes}
\begin{figure}[h!]
\centering
\includegraphics[width=0.99\hsize]{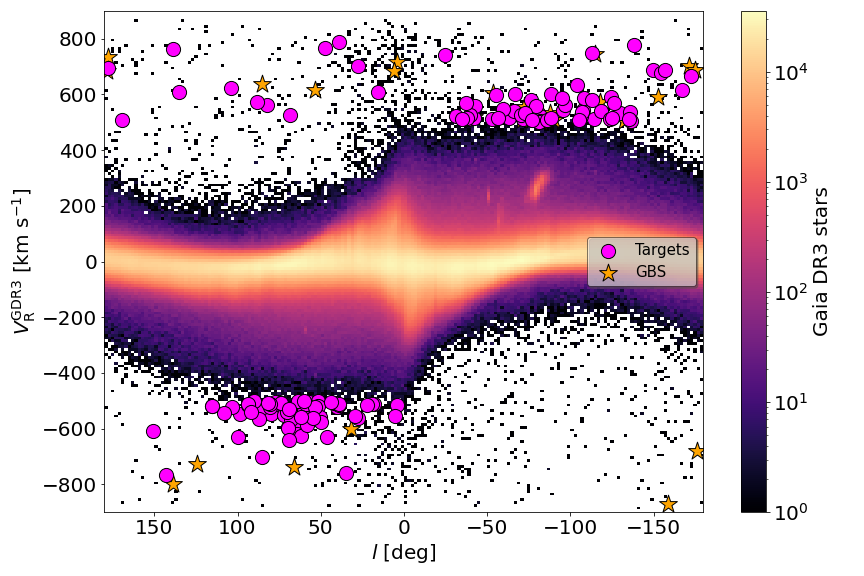}
\caption{Distribution of the 33.8 million \gdrthree\ radial velocities as a function of Galactic longitude. The \nbTargets\ SOPHIE and UVES targets and the 47 APOGEE, GALAH, RAVE, and LAMOST stars (referred to as GBS) are represented by filled magenta circles and orange stars respectively. In this and the following figures, the 2 UVES and SOPHIE targets, that are also in APOGEE or LAMOST catalogues, are only indicated with filled circles, and not with stars.}
\label{Fig:Data}
\end{figure}

Figure~\ref{Fig:Data} shows the distribution of the 33.8 million \gdrthree\ radial velocities as a function of Galactic longitude. The \nbTargets\ SOPHIE and UVES targets and the 47 APOGEE, GALAH, RAVE, and LAMOST stars (that is 49 stars minus the 2 in common with SOPHIE and UVES) are represented by filled magenta circles and orange stars respectively. The envelope of the distribution follows a sinusoidal oscillation produced by the projection effect of the Sun's Galactocentric velocity on the line of sight. The minimum and maximum of the sinusoid are populated by halo stars on retrograde orbits whose velocity vectors point in the opposite direction to the Sun's motion. A significant part of the \gdrthree\ radial velocities above and below 500 \kms\ is therefore the result of a projection effect. We return to this point in Sect.~\ref{Sect:Properties}.

\begin{table*}[t!]
\caption{Comparison of the SOPHIE and UVES radial velocities to APOGEE~DR17, LAMOST~DR7, and \citet{Matsuno2022} measurements.}
\label{Table:Surveys}
\centering
\begin{tabular}{c c c c c c}
\hline\hline
\gaia\ DR3 source\_id & Target & Spectrograph & Catalogue   & $V_\mathrm{R}$ & $\sigma_\mathrm{VR}$\\
                      &        &              &             & [\kms ]        & [\kms ] \\
\hline
1235633782930379776 & Gaia\_54 & SOPHIE       & This work   &     31.512     & 0.004 \\
1235633782930379776 &          &              & LAMOST DR7  &     33.260\tablefootmark{$\star$} & 4.430 \\ \hline
1336408284224866432 & Gaia\_2  & SOPHIE       & This work   & $-$560.471     & 0.005 \\
1336408284224866432 &          &              & Matsuno+22  & $-$560.2       &       \\ \hline
3572878053960750976 & RVS733   & UVES 	      & This work   &    521.641     & 0.5 \\
3572878053960750976 & RVS733   & UVES 	      & This work   &    520.729     & 0.5 \\
3572878053960750976 &  	       &              & APOGEE DR17 &    522.153     & 0.059 \\
\hline
\end{tabular}\\
\tablefoot{The columns V$_\mathrm{R}$ and $\sigma_\mathrm{VR}$ contain the radial velocity and radial velocity uncertainty respectively.\\
\tablefoottext{$\star$}{A correction of $+$5~$\kms$ was applied to the LAMOST radial velocity to account for the offset between LAMOST~DR7 and \gdrthree.}}
\end{table*}

\section{Measuring the radial velocities\label{Sect:Measure}}
\subsection{Cross-correlation method}
The SOPHIE pipeline incorporates a functionality to derive the radial velocity. It relies on the cross-correlation of the observed spectrum with one of the six available masks, corresponding to the spectral types M4, M5, K0, K5, G2 and F0 \citep{Bouchy2006}. The procedure is automatically launched as part of the reduction pipeline, but it can also be run off-line. The off-line mode allows to modify the radial velocity interval over which the cross-correlation function (CCF) is calculated, its sampling step as well as the mask. For each SOPHIE spectrum, we ran the off-line procedure several times iteratively, starting with the interval $[-900, +900]$~\kms\ and a step of 10~\kms\ and finishing with an interval of $\pm 20$~\kms\ around the location of the maximum of the CCF peak measured at the previous iteration and a step of 0.5~\kms. The SOPHIE masks were chosen on the basis of the temperatures of the synthetic spectra used as templates by the \gdrthree\ pipeline to measure the radial velocities and published in the catalogue in the table \verb|gaia_source|, in the field \verb|rv_template_teff|. The process converged for 62 out of the \nbTargetsSophie\ stars observed with SOPHIE. However, for Gaia\_4960, the contrast of the CCF was relatively low. Its radial velocity was confirmed by superimposing a synthetic spectrum onto the observed spectrum. For the 10 remaining stars, no peak was visible in the CCF. All but one of these stars had a \gaia\ \verb|rv_template_teff| within the range of 6000 to 7250~K. Seven had been processed using the F0 mask, and three using the G2 mask. The ten stars were reprocessed with the template-matching method described below.

\subsection{Template-matching method}
The \nbTargetsUves\ UVES stars as well as the 10 SOPHIE stars discussed above were analysed using our own template-matching code, that works in a similar way as described by \citet{Koposov2011}. The choice of template-matching was made since it works well also on broad lines, such as Balmer lines, that can otherwise give spurious results in cross-correlation methods. The UVES spectra were restricted to the wavelength range 840--880\,nm, in order to use the same spectral interval used by \gaia\ RVS. This avoids that  the comparison between our radial velocities and those of \gaia\ is perturbed by a difference in the lines used, which may otherwise result in a velocity offset. An exception was made for the star RVS725. Its spectrum in the infrared was of too low S/N to determine a radial velocity. Instead, we used the interval 600--650\,nm. The template  was a synthetic spectrum computed for each star with SYNTHE \citep[see][]{Sbordone2004,Kurucz2005} from an ATLAS9 \citep[see][]{Kurucz2005} model adopting the atmospheric parameters provided in \citet{Caffau2024, Caffau2025}. The only two exceptions were:  RVS705, that is a carbon star,  for which we adopted $\Teff\ = 4000$\,K and $\logg = 0.75$ from \gaia\ GSP-Phot parameters \citep{Andrae2023} and metallicity $+0.25$, as was used for the \gaia\ RVS template; the variable and active star RVS724, for which we adopted $\Teff = 4750$\,K and $\logg = 2.00$, close to the spectrophotometric \gaia\ values, and for the metallicity we adopted $-0.25$, from our visual estimate. Reliable radial velocities were measured for all UVES stars and for 7 out of the 10 SOPHIE stars. For the remaining three stars, Gaia\_44, Gaia\_1137 and Gaia\_820, we were nevertheless able to rule out the high radial velocities provided in the \gdrthree\ catalogue. Gaia\_44 shows a double emission feature in H$\alpha$ with velocities of $-82$ and $+188$~\kms\ and a possible absorption at $+101$~\kms. Its S/N is too low to use other spectral lines. Gaia\_1137 exhibits a strong emission in the core of H$\beta$ and the locations of the Balmer lines are consistent with a radial velocity close to 0~\kms. No other spectral lines are visible, suggesting a hot star (consistent with the extended wings of the Balmer lines) and/or possibly very metal-poor star. Gaia\_820 has no metallic lines either, but H$\alpha$ and H$\beta$ exhibit very large cores (suggestive of an SB2 binary) with small velocity shifts inconsistent with the radial velocity published in \gdrthree\ of 614~\kms. We note that Gaia\_4984 also shows emission in the core of H$\beta$, but it was nonetheless possible to measure a velocity of 12.4~\kms, which is incompatible with the 778.6 \kms\ published in \gdrthree.

\subsection{Radial velocity measurements}
The SOPHIE and UVES radial velocities of our targets are provided in Table~A.1, which is available online and is described in Appendix~\ref{App:Targets}. The statistical uncertainty on the UVES radial velocities ranges from 1\,\ms\ to 30\,\ms , because the S/N of the spectra is very high. However, the total uncertainty is dominated by the systematic error, which itself is mainly due to the decentring of the star on the slit. To account for this, a conservative value of 0.5 \kms\ (corresponding to a decentring of approximately 0\farcs{05} on the slit) was added in quadrature to the statistical uncertainty. The uncertainty on the radial velocities measured from the SOPHIE spectra is typically of the order of a few \ms.

We utilised the eight stars for which both SOPHIE and UVES spectra are available to assess the consistency between the two employed methods: cross-correlation and template-matching. The radial velocities derived from the two techniques are in good agreement, exhibiting a median difference of $-$0.89~\kms\ and a Median Absolute Deviation (MAD) of the differences of 0.21 \kms.

As mentioned in Sect.~\ref{Sect:Surveys}, two of our targets are also part of the APOGEE~DR17 \citep{Abdurrouf2022} and LAMOST~DR7 \citep[see for the survey and pipeline description,][]{Zhao2012, Deng2012, Luo2015} catalogues respectively. Moreover, the SOPHIE target Gaia\_2 has also been observed by \citet{Matsuno2022}. The radial velocity measurements for these three stars are shown in Table~\ref{Table:Surveys}. A correction of 5~\kms\ was added to the LAMOST radial velocity to account for the offset between LAMOST~DR7 and \gdrthree\ \citep[see Sect. 6.2 in][]{Katz2023}. After correction, all values agree to within less than 2~\kms.

Thirty one single line targets were observed several times with SOPHIE and/or UVES. Three of them, namely RVS745, RVS711 and RVS712, show a difference between the smallest and largest radial velocity measurements of over 3~\kms, that is 3, 5.4 and 12.8 \kms\ respectively. These differences are much larger than the formal uncertainties on the radial velocities which are of the order of a few \ms\ to a few hundreds of \ms. There is therefore a strong probability that these stars belong to binary systems or, possibly, are variable. However, further observations are needed to confirm this. The mean difference between the smallest and largest radial velocity values for the other 28 stars is 0.7 \kms. In Sects.~\ref{Sect:Confirmed}, \ref{Sect:OutliersRate} and \ref{Sect:Properties}, when a star has several observations, we use the mean value of the measurements of its radial velocity.

Two targets are identified as variable in the literature. The star Gaia\_16 is classified as an RR Lyrae by \citet{Drake2013} and RVS708 is classified as a long period variable by \citet{Heinze2018}. We collected only a single SOPHIE spectrum for the first and a single UVES spectrum for the second, but they yield radial velocities in agreement with \gdrthree\ values within 2-3 \kms.

Finally, Gaia\_432 turned out to be a double-line spectroscopic binary (SB2). The velocities of the two components were determined by fitting double Voigt profiles to the CCFs. The measures are presented in Table~\ref{Table:SB2}. The values provided in Table~A.1 are the average of the two components, for each observation. They vary between 568.222 and 568.508 \kms. This is very close to the value published in \gdrthree, that is 568.443 \kms, indicating that the Gaia pipeline measured the mean location of the two components.

\begin{table*}[h!]
\caption{Velocities of the two components of the SB2 star Gaia\_432.}
\label{Table:SB2}
\centering
\begin{tabular}{c c c c c c}
\hline\hline
\gaia\ DR3 source\_id & MJD & $V_\mathrm{R}$ comp1 & $\sigma_\mathrm{VR}$ comp1 & $V_\mathrm{R}$ comp2 & $\sigma_\mathrm{VR}$ comp2\\
 & [d] & [\kms ] & [\kms ] & [\kms ] & [\kms ] \\
\hline
5757799278315703296 & 59962.016701 & 546.058 & 0.320 & 590.615 & 0.307 \\
5757799278315703296 & 59962.106678 & 546.605 & 0.290 & 590.412 & 0.282 \\
5757799278315703296 & 59962.970023 & 547.403 & 0.739 & 589.041 & 0.533 \\
5757799278315703296 & 59963.089873 & 548.233 & 0.499 & 588.698 & 0.225 \\
\hline
\end{tabular}\\
\tablefoot{The Modified Julian Date (MJD) corresponds to the start of the exposure.}
\end{table*}

\begin{figure}[h!]
\centering
\includegraphics[width=0.99\hsize]{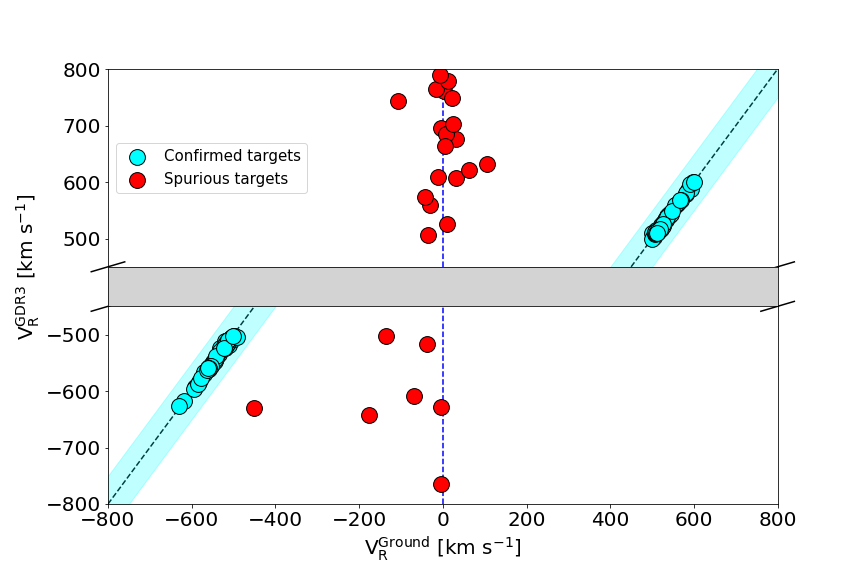}
\caption{\gdrthree\ \texttt{radial\_velocity} field versus SOPHIE and UVES radial velocities. The 104 confirmed stars are represented by filled cyan circles, the 27 spurious measures with a ground-based radial velocity by filled red circles and the shaded cyan area indicates the validity zone of $\pm 50$~\kms\ around the main diagonal. For the sake of clarity, the central part of the figure, which was empty, has been omitted and has been replaced by a narrow grey band.}
\label{Fig:Confirmed}
\end{figure}

\section{Confirmed HRV and spurious measures\label{Sect:Confirmed}}
By selection, our targets all have very large radial velocities in \gdrthree. This makes it much easier to confirm or refute these measurements. Figure~\ref{Fig:Confirmed} compares the \gdrthree\ \verb|radial_velocity| field (ordinates) to the radial velocities measured with SOPHIE and UVES (abscissas). The stars mostly split in two groups. The first group is distributed along the main diagonal and is made of stars with consistent measurements between \gaia\ and the ground. The second group is made of stars with ground-based velocities in the range roughly $\pm 200$~\kms, which are totally incompatible with the values published in \gdrthree. As a quantitative criterion, we considered that the \gdrthree\ radial velocities were confirmed if they were consistent with the ground-based velocities to within $\pm 50$~\kms, and otherwise, they were deemed disproved. We deliberately chose a high value of 50 \kms\, to take into account the possible variability or multiplicity of our targets. In practice, the largest absolute residual in our confirmed star sample is 11.2~\kms. In Fig.~\ref{Fig:Confirmed}, the \gdrthree\ confirmed stars are represented by filled cyan circles, the spurious measures by filled red circles and the shaded cyan area indicates the validity zone of $\pm 50$~\kms\ around the main diagonal. There is one star that falls in between the two well separated groups. Gaia\_6151 has a SOPHIE velocity of $-450.1$~\kms\ and a \gdrthree\ \verb|radial_velocity| of $-630.7$~\kms. Its radial velocity residual of $\sim 180$~\kms\ classifies it as a spurious measure (according to our criterion), but at the same time its SOPHIE velocity is that of a high radial velocity star. We therefore cannot rule out the possibility that this star could belong to a tight binary system, with a semi-amplitude of the orbital velocity of the order of 200 \kms. By applying the criterion presented above, the \gdrthree\ \verb|radial_velocity| of 104 of our \nbTargets\ targets are confirmed and 27 refuted. To these should be added the three stars without a measured SOPHIE or UVES velocity, but with H$\alpha$ or H$\beta$ locations incompatible with their \gdrthree\ radial velocities (see Sect.~\ref{Sect:Measure}), for a total of 30 spurious measurements. The median and standard deviation of the velocity residuals of the confirmed sample are 0.4 and 3~\kms\ respectively.

\section{Outlier rate\label{Sect:OutliersRate}}
\subsection{Outlier rate in \gdrthree\ HRV stars}
\begin{figure}[h!]
\centering
\includegraphics[width=0.99\hsize]{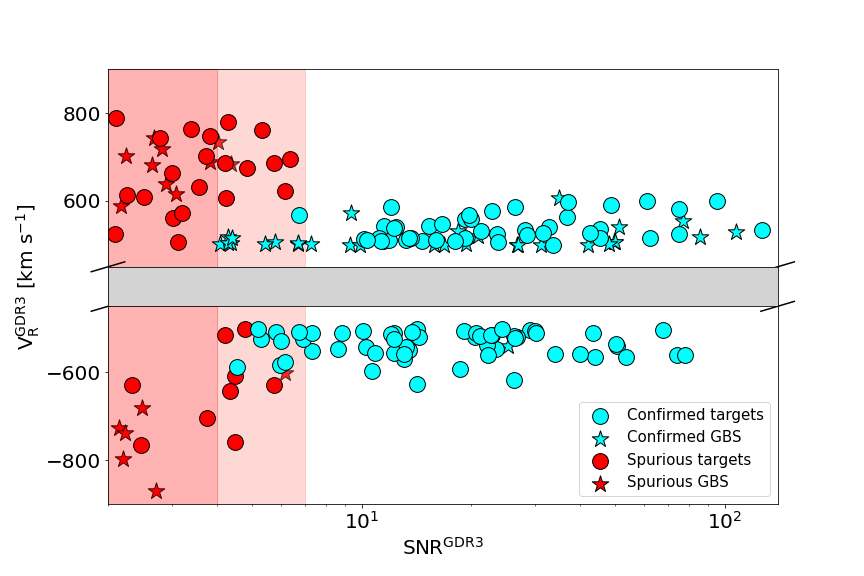}
\caption{\gdrthree\ radial velocities as a function of the \gdrthree\ S/N, for our \nbTargets\ targets (filled circles) and for the 47 APOGEE, GALAH, LAMOST and RAVE stars (stars). The confirmed \gdrthree\ values are coloured in cyan and the spurious measures are coloured in red. As in Fig.~\ref{Fig:Confirmed}, the central part of the figure has been omitted and replaced by a narrow grey band.}
\label{Fig:Outliers1}
\end{figure}

Figure~\ref{Fig:Outliers1} shows the \gdrthree\ radial velocities (field \verb|radial_velocity|) as a function of the \gdrthree\ S/N (field \verb|rv_expected_sig_to_noise|) for our \nbTargets\ targets (filled circles) as well as for the 47 APOGEE, GALAH, LAMOST and RAVE stars (stars). The confirmed values are coloured in cyan and the spurious measures are coloured in red. As the figure shows, for the stars in our sample, the S/N correlates strongly with the reliability of the \gdrthree\ radial velocities. For the S/N in the range 2 to 4 (red shaded area), all values are spurious. From 4 to 7 (salmon shaded area), there is a mix of spurious measures and valid velocities. Beyond S/N 7 (white area), the \gdrthree\ radial velocities of all our targets are confirmed. Of course, the 181 stars considered here represent just over 10\% of the 1544 \gdrthree\ HRV. The boundaries between the spurious and confirmed values for the full set are probably not strictly those derived from this limited sample. Moreover, it can happen that a wrong choice of template in the \gdrthree\ pipeline or the contamination by a brighter neighbouring star do spoil the radial velocity of a high S/N star. In any case, Figure~\ref{Fig:Outliers1} provides guidance on the reliability of the \gdrthree\ HRV stars. For a very reliable sample of fast-moving stars, one should probably discard sources with S/N below 7, or even slightly higher.

As shown in Figure~\ref{Fig:Outliers1}, within the HRV sample considered here, the S/N appears to be the primary factor influencing the reliability of the \gdrthree\ radial velocities. However, it is reasonable to question whether this is a multifactorial issue, and whether contamination by nearby bright stars also plays a significant role. As described in Section 4.3.3 of \citet{Katz2023}, prior to the release of \gdrthree, all stars with an absolute value of the radial velocity exceeding 200 \kms\ and having a brighter neighbour (in the $G$-band) within 10~arcseconds were examined. Those identified as contaminated were excluded from the published catalogue. To assess whether the \gdrthree\ filtering may have been overly permissive, we re-examined the vicinity of the sample stars exhibiting spurious \gdrthree\ radial velocities, using an extended search radius of 15 arcseconds. Among the 46 stars in question, only five (11\%) have a brighter neighbour in the $G$-band, and only two appear both close enough and bright enough to possibly be contaminated.

\begin{figure}[h!]
\centering
\includegraphics[width=0.99\hsize]{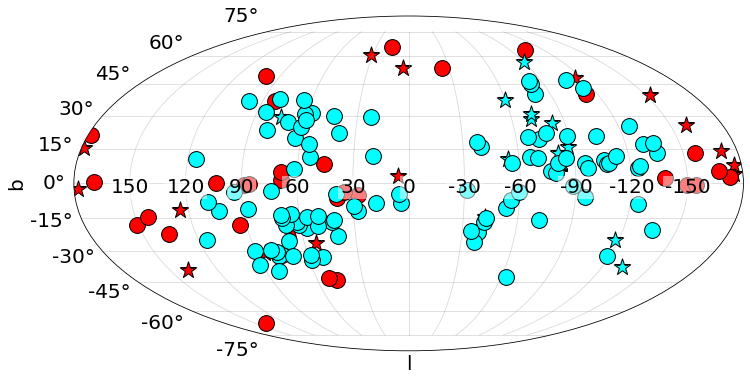}
\caption{Distribution in Galactic longitude ($l$) and latitude ($b$) of the 181 HRV stars. The figure uses a Mollweide projection. As in Fig.~\ref{Fig:Outliers1}, our targets are represented by filled circles and the GBS by filled star symbols. Stars with confirmed \gdrthree\ radial velocities are displayed in cyan, while those with spurious measurements are indicated in red.}
\label{Fig:lb}
\end{figure}

An alternative approach to this question is to consider the spatial distribution of stars with spurious versus confirmed \gdrthree\ radial velocities. Figure~\ref{Fig:lb} presents the Galactic longitudes ($l$) and latitudes ($b$) of the 181 stars in our sample. The proportion of stars located at low Galactic latitudes (defined here as $-20 < b < 20$~deg) is similar in both groups: 61\% for the confirmed sample and 57\% for the spurious one. The latter are therefore not disproportionately clustered in the dense regions of the Galactic disc compared to the confirmed stars. Taken together, these findings support the conclusion that the primary cause of spurious \gdrthree\ radial velocities among HRV stars is low S/N. Nonetheless, when analysing individual stars, it remains prudent to check for potential contamination, especially in cases where the absolute value of the \gdrthree\ radial velocity is less than 200~\kms, as such stars were not subject to neighbour inspection prior to the catalogue’s release.

\subsection{Outlier rate in the \gdrthree\ catalogue}
Figure~\ref{Fig:Outliers1} could convey the wrong impression that most \gdrthree\ radial velocities derived from RVS spectra with S/N below 4 are false. This is fortunately not the case. As we saw in the introduction, very low S/N spectra could give rise to cross-correlation functions (CCF) that are so noisy that the noise peaks can exceed the true CC peak. These spurious secondary peaks are more or less uniformly distributed over the velocity range probed by the CCF, that is $-$1000 to $+$1000 \kms\ for \gdrthree. The distribution of spurious radial velocities resulting from very low S/N spectra is therefore roughly flat. In contrast, the radial velocities of Milky Way stars published in the \gdrthree\ catalogue peak at around 0 \kms\ and decrease rapidly at lower and higher velocities. The rate of contamination is therefore maximal in the outer wings (below $-$500 and above $+$500 \kms), as they are very sparsely populated. The situation improves for more central radial velocities, with the flat distribution of outliers representing an increasingly small fraction of the radial velocities.

\begin{figure}[h!]
\centering
\includegraphics[width=0.99\hsize]{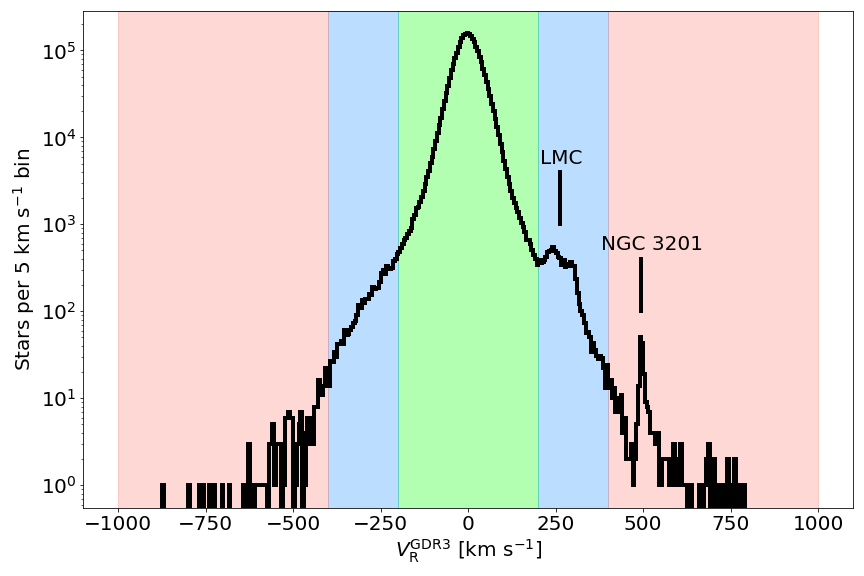}
\caption{Distribution of the \gdrthree\ radial velocities of the sample of 2.7 million stars used to assess the rate of outliers as a function of S/N and for three different intervals of absolute values of the radial velocities. The three intervals are materialised in green ([0, 200) \kms), blue ([200, 400) \kms) and salmon ([400, 1000) \kms) respectively. We note that the Large Magellanic Cloud (LMC) and the Globular Cluster NGC~3201 both produce a peak visible on top of the distribution of Milky Way stars.}
\label{Fig:Outliers2}
\end{figure}

In order to quantify the outlier rate in different radial velocity ranges, we cross-matched the APOGEE DR17, GALAH DR3, GES DR3, LAMOST DR7 and RAVE DR6 catalogues with the \gdrthree\ data. The sample obtained is made of around 2.7~million stars with both \gdrthree\ radial velocities and velocities measured from the ground, including our 104 confirmed targets. Figure~\ref{Fig:Outliers2} shows the distribution of the \gdrthree\ radial velocities of these stars. The sample was then divided into three groups, based on the absolute value of \gdrthree\ radial velocities: that is [0, 200), [200, 400) and [400, 1000) \kms. In Fig.~\ref{Fig:Outliers2}, the three groups correspond to the green, blue and salmon areas respectively.

\begin{figure}[h!]
\centering
\includegraphics[width=0.99\hsize]{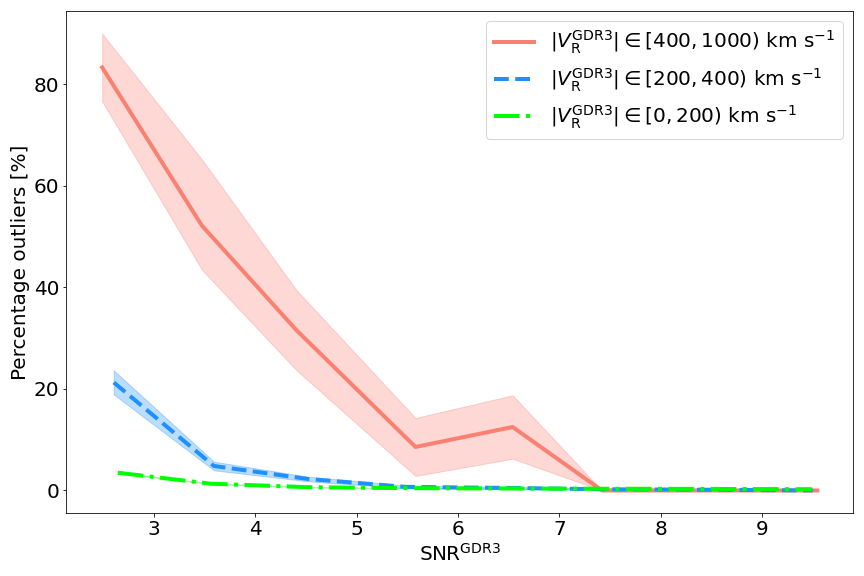}
\caption{Percentage of outliers as a function of S/N and for the three velocity groups: [0, 200) \kms\ (dashed-dotted green line), [200, 400) \kms\ (dashed blue line), and [400, 1000) \kms (solid salmon-coloured line). The shaded areas materialise the 1$\sigma$ error bars.}
\label{Fig:Outliers3}
\end{figure}

In each of the three groups, the stars were sorted into bins of 1 in S/N, and in each bin the spurious measures were counted. We used the same validity criterion as in Sect.~\ref{Sect:Confirmed}. A \gdrthree\ radial velocity is considered spurious if the absolute value of the radial velocity residual (\gaia\ minus ground) is larger than 50 \kms. The uncertainties were estimated via bootstrap resampling. For each bin, 1000 iterations were performed by randomly drawing $N$ absolute radial velocity residuals (with replacement), from the original set of $N$. Figure~\ref{Fig:Outliers3} shows the percentage of outliers as a function of S/N and for the three velocity groups. As expected, the least populated velocity range $(-1000, -400] \cup [400, 1000)$~\kms exhibits the highest rates of spurious measurements, while the most populated range $(-200, 200)$~\kms shows the smallest rates. For example, in the S/N bin [2, 3), the percentage of outliers is 83\% for \gdrthree\ \texttt{|radial\_velocity|} in [400, 1000) \kms, 21\% for [200, 400) \kms\ and 3.5\% for [0, 200) \kms. These results are applicable exclusively to \gdrthree. In \gdrfour, new techniques should be employed to improve the identification and exclusion of the spurious measurements.

As discussed in Sect.~\ref{Sect:Confirmed}, the use of a fairly high threshold of 50 \kms\ is motivated by the possible multiplicity or variability of the sources. In order to measure the impact of the choice of the threshold on the final results, we repeated the study with a threshold of 25 \kms. In the high radial velocity range ([400, 1000) \kms), lowering the threshold has little effect. Indeed, \gdrthree\ velocities are so large that it is easy to distinguish between valid and spurious ones (as we saw in Sect.~\ref{Sect:Confirmed}). On the other hand, as the \gdrthree\ velocities decrease, the ambiguity between valid and erroneous measures increases, and the choice of threshold weighs more heavily on the percentage of outliers. If we take the example of the S/N bin [2, 3) with a threshold of 25 \kms, we obtain the following outlier percentages: 83\% for [400, 1000) \kms, 26\% for [200, 400) \kms\ and 11.5\% for [0, 200) \kms.

Figure~\ref{Fig:Outliers3} shows that \gdrthree\ S/N can be used to mitigate the rate of outliers in \gdrthree\ radial velocities. The threshold should be adapted to each specific science case, based on its tolerance to outliers.

\section{HRV properties\label{Sect:Properties}}
In this section, we return to the HRV stars and examine their properties within the Milky Way.

\subsection{Selection effect\label{Sect:Selection}}
\begin{figure}[h!]
\centering
\includegraphics[width=0.99\hsize]{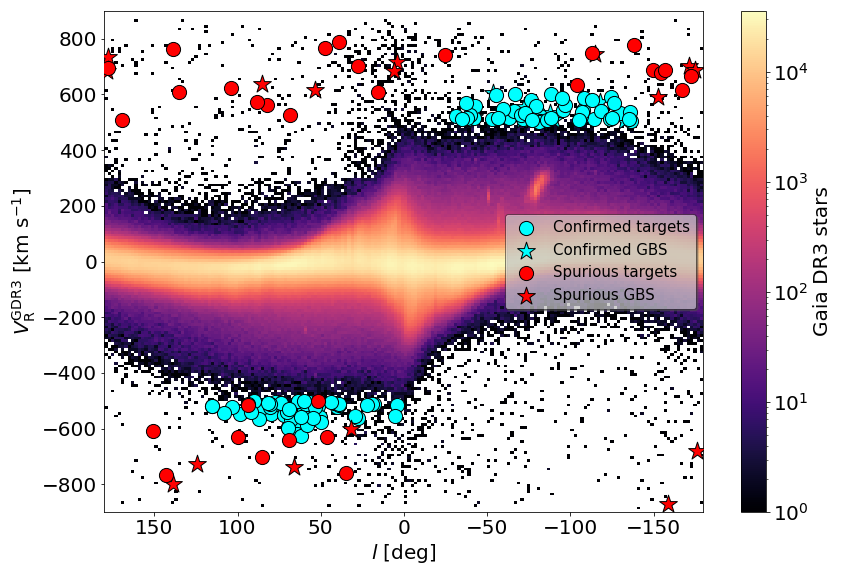}
\caption{Same as Fig.~\ref{Fig:Data}, except for the colour coding of the symbols which distinguishes the confirmed radial velocities (cyan) and the spurious measures (red).}
\label{Fig:Properties1}
\end{figure}

Figure~\ref{Fig:Properties1} is similar to Fig.~\ref{Fig:Data}. The distribution of the \gdrthree\ radial velocities as a function of Galactic longitude is shown as a density map, while the \nbTargets\ SOPHIE-UVES targets and the 47 APOGEE, GALAH, RAVE, and LAMOST stars are represented by filled circles and stars respectively. What is different here from Fig.~\ref{Fig:Data} is the colour coding which distinguishes the confirmed radial velocities (cyan) from the spurious measures (red). The confirmed \gdrthree\ radial velocities cluster at the outer edges of the distribution, that is at positive velocities for negative longitudes and at negative velocities for positive longitudes. These outer edges are populated by stars on retrograde orbits, whose velocity vector points in the opposite direction to the Sun's motion, making it 'easier' to reach large radial velocities measured with respect to the Solar System barycentre reference frame. As a result, when one selects HRV stars from the \gdrthree\ catalogue, one is primarily selecting stars in retrograde orbits.

\begin{figure*}[h!]
\centering
\includegraphics[width=0.49\hsize]{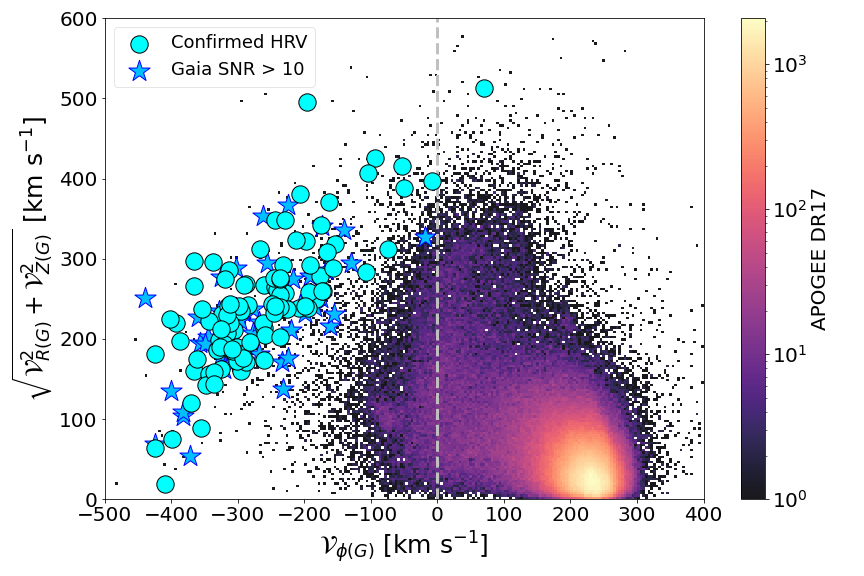}
\includegraphics[width=0.49\hsize]{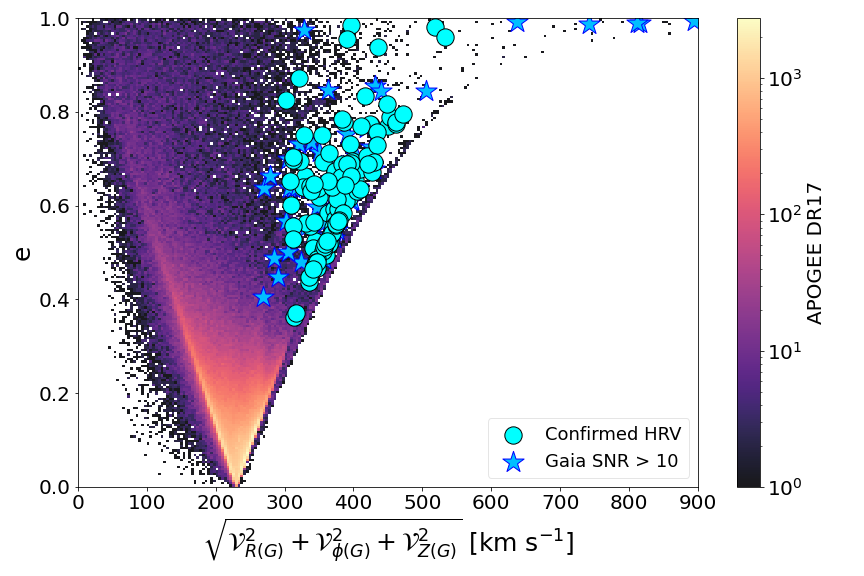}
\includegraphics[width=0.49\hsize]{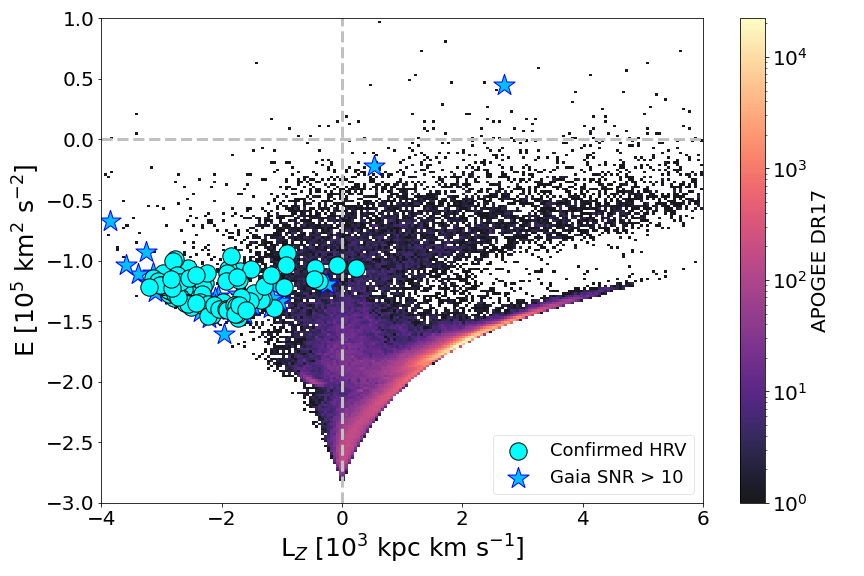}
\includegraphics[width=0.49\hsize]{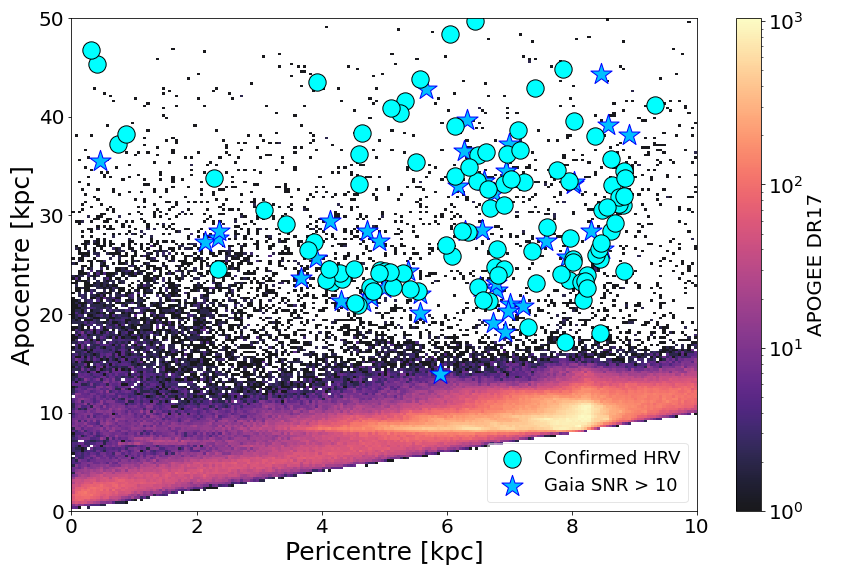}
\caption{Location of the 161 clean HRV sample stars, in four different diagrams: (top left) Toomre diagram, (top right) eccentricity versus total velocity, (bottom left) energy versus vertical component of the angular momentum, and (bottom right) maximum versus minimum distance to the Galactic centre. The 108 confirmed HRV stars are represented by cyan filled circles and the 53 \gdrthree\ stars with S/N > 10, by blue stars. The density maps are made of APOGEE DR17 stars.}
\label{Fig:Properties2}
\end{figure*}

\subsection{Kinematic and dynamic properties of the HRV stars\label{Sect:Dynamic}}
In this section, we examine the kinematic and dynamic properties of a clean sample of HRV stars. It is made of two groups. The first contains the 108 \gdrthree\ HRV stars with a ground-based confirmation (either from our UVES-SOPHIE programme or from APOGEE, GALAH, and RAVE), and which satisfy the astrometric quality criteria of \verb|RUWE|~<~1.4 and \verb|parallax_over_error|~>~5. The second comprises the \gdrthree\ HRV stars that lack ground-based confirmation, but possess a S/N greater than 10, and satisfy the same astrometric quality criteria. This second group consists of 53 stars. The orbits of the 161 stars were integrated using a leap-frog algorithm and the axisymmetric Galactic potential 'PII' of \citet{Pouliasis2017}. This potential includes both a thin and a thick disc as well as a spherical dark matter halo. For the astrometric data, we employed the \gedrthree\ positions and proper motions \citep{GaiaCollaborationBrown2021, Lindegren2021astrometry}, along with the Bayesian photogeometric distances derived by \citet{BailerJones2021}. As a consistency check, we compared the photogeometric distances with those derived directly from the parallaxes, using the relation $d = 1000 / (\varpi - \varpi_0)$, where $d$ is the distance in parsecs, and $\varpi$ and $\varpi_0$ the parallax and parallax zero-point \citep{Lindegren2021bias}, both in milliarcseconds. The median absolute difference is 1.8\%, with the largest difference obtained for the star \gdrthree\ 3466996351219554944, whose bayesian distance (5597~pc) is 15\% smaller than its distance derived directly from its corrected parallax (6580~pc). We used the distance Sun-Galactic centre $R_\odot = 8.277$~kpc \citep{Gravity2022} and the vertical distance Sun-Galactic mid-plane $Z_\odot = 20.8$~pc \citep{BennettBovy2019}. For the solar velocity with respect to the Galactocentric cylindrical reference frame, we adopted the same values as in \citet{Drimmel2023}, that is $\mathcal{V}^\odot_\mathrm{R(G)} = 9.3$~\kms, $\mathcal{V}^\odot_\mathrm{\phi(G)} = 251.5$~\kms and $\mathcal{V}^\odot_\mathrm{Z(G)} = 8.59$~\kms. The Galactocentric cylindrical reference frame used here is left-handed, that is the radius ($R$) is defined positive in the direction away from the Galactic centre, the azimuth ($\phi$) is defined positive in the direction of the Galactic rotation, and the vertical coordinate ($Z$) is defined positive in the direction of the Galactic north pole. It should be emphasised that, in this section as well as in Fig.~\ref{Fig:Properties2}, we use $\mathcal{V}_\mathrm{R(G)}$ to refer to the radial velocity in the Galactocentric cylindrical reference frame (and $\mathcal{V}_\mathrm{\phi(G)}$ and $\mathcal{V}_\mathrm{Z(G)}$ for the azimuthal and vertical components), while in the rest of the paper, we use $V_\mathrm{R}$ to refer to the radial velocity in the solar barycentric frame.

Figure~\ref{Fig:Properties2} shows the location of the 161 clean HRV sample stars, in four different diagrams: on the top left,  Toomre diagram, on the top right, eccentricity versus total velocity, on the bottom left, energy versus vertical component of the angular momentum, and on the bottom right, maximum versus minimum distance to the Galactic centre. The 2D histograms of APOGEE DR17 stars provide references for the typical distributions of the stars in the disc, bulge and inner halo. In the $E$--$L_\mathrm{Z}$ diagram, the elongated structure at positive $L_\mathrm{Z}$, and situated at a higher energy than the Milky Way disc, corresponds to the Sagittarius Dwarf Spheroidal Galaxy \citep{Ibata1994}.

Figure~\ref{Fig:Properties2} top left and bottom left panels show that the vast majority of the HRV stars are on retrograde orbits. The median azimuthal velocity ($\mathcal{V}_\mathrm{\phi(G)}$) of the sample is $-284$~\kms\, with 92.5\% of the stars exhibiting an azimuthal velocity less than $-$100~\kms. The median value of the vertical component of the angular momentum ($L_\mathrm{Z}$) is $-2.2 \times 10^3$ kpc~\kms. As discussed in Sect.~\ref{Sect:Selection} above, the selection of HRV stars inherently favours those moving in a direction opposite to that of the Sun, and therefore on retrograde orbits. For the most part, the HRV stars exhibit total velocities within the range 300 to 450 \kms (top right panel) and remain gravitationally bound to the Milky Way, as indicated by their negative energies (bottom left panel). The stars mostly follow eccentric orbits, with almost 99\% having an eccentricity ($e$) larger than 0.4 (top right panel). The median values of the pericentres and apocentres of the HRV stars are 6.6 and 28.4~kpc, respectively (bottom right panel). Six stars have pericentres located within 1~kpc of the Galactic centre. All six follow highly eccentric orbits, with eccentricities exceeding 0.95.

As shown in the bottom left panel of Fig.~\ref{Fig:Properties2}, the bulk of the HRV stars exhibit energies ($E$) in the range $-$1.5 to $-1.0 \times 10^5$~km$^2$~s$^{-2}$ and vertical components of the angular momentum ($L_\mathrm{Z}$) between $-$4 and $-1 \times 10^3$~kpc~\kms. Several structures associated with past merger events have been identified in this area, in particular: Sequoia \citep{Myeong2018, Myeong2019, Matsuno2019, Matsuno2022}, Arjuna and I'itoi \citep{Naidu2020}, Antaeus \citep{Oria2022}, ED-2 and ED-3 \citep{Dodd2023}. Interestingly, Gaia\_432, the SB2 system discussed in Sect.~\ref{Sect:Measure}, has been identified as a member of Antaeus by \citet{Oria2022}. The lowest-energy HRV are situated near the upper boundary of the Thamnos 1 and 2 groups \citep{Koppelman2019}, which are more tightly bound gravitationally to the Milky Way than the previously mentioned structures. It is likely that the majority of the HRV stars discussed here were accreted.

Finally, four stars exhibit positive energies: \gdrthree\ 1061822400696509696, 2020578982307467776, 3017364441287016320 and 5943642753740584320. The latter star (\gdrthree\ 5943...4320) is visible in the $E$-$L_\mathrm{Z}$ diagram, while the remaining three, which lie outside the boundaries of the figure, have energies in the range $1.3 \times 10^5$ to $2 \times 10^5$~km$^2$~s$^{-2}$ and vertical components of the angular momentum between $4.6 \times 10^3$ and $6.6 \times 10^3$~kpc~\kms. All four stars belong to the sample of 53 stars without ground-based confirmation of their radial velocity, but an S/N greater than 10. In examining indicators of data quality, we note that the star \gdrthree\ 3017...6320 exhibits an \verb|astrometric_gof_al| value of –7.38, indicative of a poor fit between the astrometric solution and the along-scan observations, possibly caused by binarity. Additionally, its \gdrthree\ radial velocity is based on only five valid transits (\verb|rv_nb_transits|), in contrast to the median value of 18 valid RVS transits per star in \gdrthree. Lastly, according to SIMBAD\footnote{https://simbad.cds.unistra.fr/simbad/} (CDS, Strasbourg), the star is classified as an Orion Variable. Taken together, these elements suggest that the velocity and orbital parameters derived here for \gdrthree\ 3017...6320 may be unreliable. Conversely, the three other stars exhibit satisfactory \verb|astrometric_gof_al| values, ranging from –1.1 to 2.9. Their \gdrthree\ radial velocities were derived based on 19 to 26 valid transits. Furthermore, the ratios of their parallax corrected for the parallax zero-point \citep[$\varpi - \varpi_0$;][]{Lindegren2021bias} and proper-motions ($\mu_\alpha$, $\mu_\delta$) over the corresponding errors ($\sigma_\varpi$, $\sigma_{\mu_\alpha}$, $\sigma_{\mu_\delta}$) are all relatively large, that is $(\varpi - \varpi_0) / \sigma_\varpi \in \{357, 22, 12\}$, $\mu_\alpha / \sigma_{\mu_\alpha} = \{799, 91, 136\}$, and $\mu_\delta / \sigma_{\mu_\delta} = \{284, 267, 290\}$ for Gaia DR3 1061...9696, 2020...7776, and 5943...4320, respectively. However, the radial velocities of \gdrthree\ 2020...7776 and 5943...4320 were both determined by the \gaia\ spectroscopic pipeline using a default template, which is employed when the stellar parameters are unknown. This may introduce a bias in the radial velocity measurements, if the true stellar parameters differ significantly from those of the default template \citep{Katz2023}. For the third star, \gdrthree\ 1061...9696, the template used had an effective temperature of 3400~K. In this case as well, there is a risk of bias arising from the possible presence of molecular features in the spectrum. The three stars are visible in the eccentricity-total velocity diagram, displaying eccentricities greater than 0.98 and total velocities ranging from 742 to 895 \kms. They are all on prograde orbits, with azimuthal velocities between 570 and 885 \kms. Their positive total energies suggest that these stars are gravitationally unbound from the Milky Way, and their total Galactocentric velocities classify them as HVSs \citep{Carballo-Bello2025}. However, their total Galactocentric velocities are primarily driven by the unconfirmed \gdrthree\ radial velocity measurements and as discussed above the three stars present a risk of radial velocity bias induced by template mismatch. Future follow-up ground-based spectroscopic observations are needed to confirm or refute their HVS status.

\begin{figure}
\centering
    \resizebox{8.cm}{!}{\includegraphics{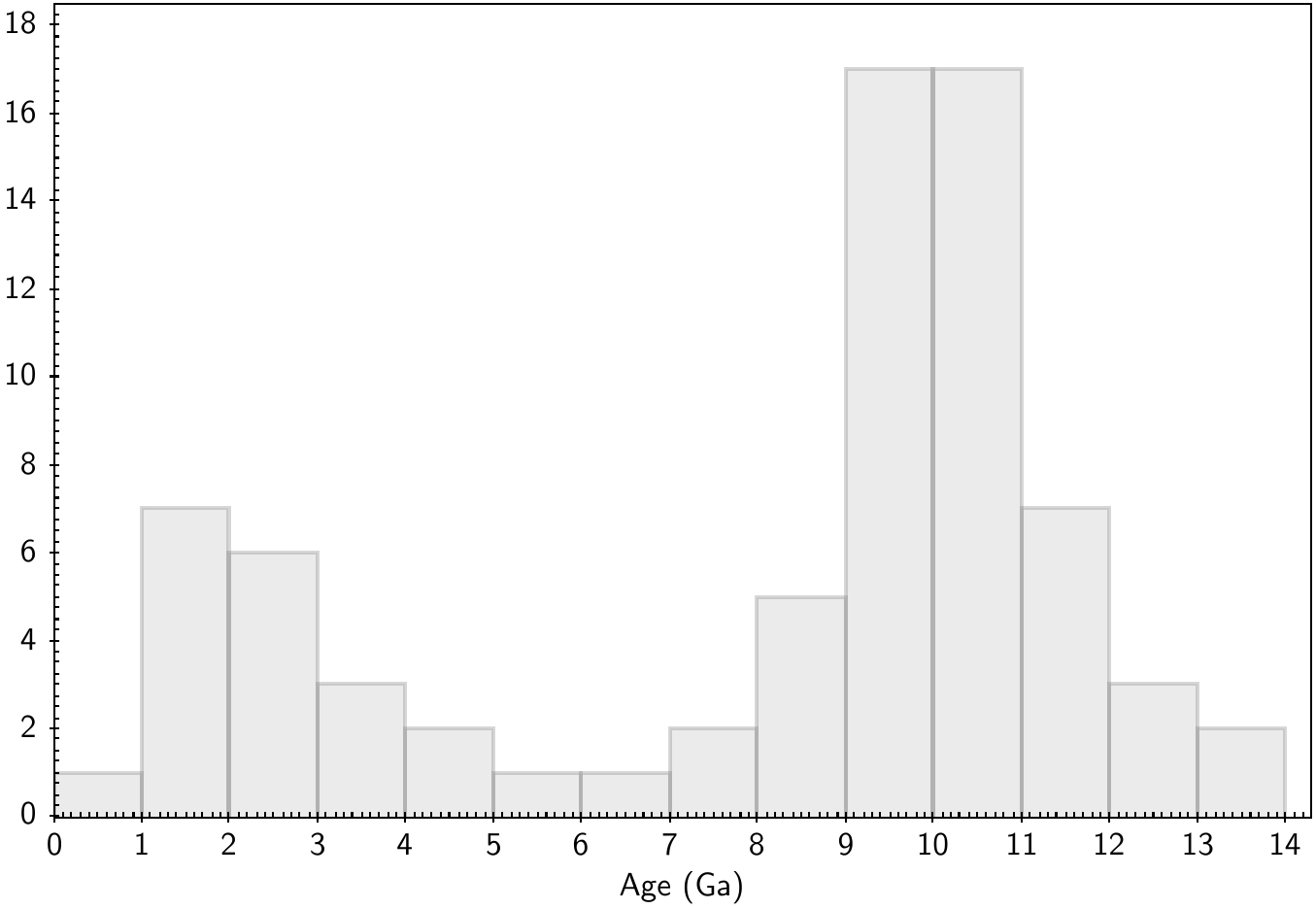}}
    \caption{Histogram of the ages of the 74 stars with the most reliable ages.}
    \label{fig:age_histo}
    \centering
\end{figure}

\subsection{Age properties of the HRV stars}
We used the luminosities derived from the \gaia\ $G$ magnitudes and parallaxes, and our derived effective temperatures and metallicities as input to SPInS \citep{LebretonReese2020,spins2020} to derive the ages and masses of our sample of stars. As grid of input stellar evolutionary tracks we used the $\alpha$ enhanced set of BaSTI \citep{Pietrinferni2021}. The SPInS inference provides an output metallicity which generally differs from the input metallicity. We retained as robust age estimates only those for which the absolute difference between input and output metallicity is less than 0.2\,dex. This sample consists of 74 turnoff and giant stars. For this sample the uncertainty on the age ranges from 0.1\,Ga to 3.8\,Ga, with lower uncertainties observed for subgiants and increasing uncertainties for more luminous stars.

In Fig.~\ref{fig:age_histo},  we show the histogram of the ages of this sample of stars. A clear bimodality is present. The main group consists of stars older than 7 -- 8\,Ga. This is consistent with the expected age of the bulk of the accreted halo population \citep[see e.g.][]{Kruijssen2020}, including some of the retrograde structures, such as for example Sequoia \citep{Myeong2019} or Antaeus \citep{Oria2022}, which exhibit dynamical similarities with the HRV stars (as discussed in Sect.~\ref{Sect:Dynamic}). The second group is made of stars younger than 6\,Ga. These stars could be blue stragglers or evolved blue stragglers \citep{Jofre2023, Cerqui2023}. However, as discussed in \citet{Caffau2024} some could be truly young metal-poor stars formed in dwarf galaxies at the time they entered in interaction with the Milky Way hot gas, and subsequently accreted by the Milky Way. The latter hypothesis is more likely for the most massive stars, those that exceed 1.6~$M_\odot$ and that can hardly be formed by the fusion of two old turnoff stars. The issue is discussed at length in \citet{Bonifacio2024}. Since the mass of an old turnoff (TO) metal-poor star is about 0.8~$M_\odot$ a fusion of two such stars, assuming no loss of mass in the fusion process, could form a Blue Straggler Star (BSS) of at most 1.6~$M_\odot$. Slightly larger masses of the two TO stars could be invoked if the parent population  were as young as 6\,Ga, or younger, in which case the mass of the single TO star could rise up to 0.9~$M_\odot$. However, 6\,Ga is already pretty young for a metal-poor star and in any case the mass of the BSS would not exceed 1.8~$M_\odot$. To achieve more massive BSS it is necessary to invoke the fusion of three stars in a triple system \citep{Meyer1980}. This process certainly occurs, as testified by the BSS that exceed twice the mass of the TO stars in the open cluster M\,67 \citep{Milone1992} and globular cluster NGC\,6397 \citep{Saffer2002}. However these events are very rare. \citet{Bonifacio2024} estimate that only 0.3\% of the BSS may have been formed in this way. Moreover, fusion is probably not the preferred mechanism for BSS formation, while accretion seems favoured \citep{Preston2000}. In the Globular Cluster NGC\,6541, \citet{Fiorentino2014} derived masses in the range 1.0--1.1\,$M_\odot$. One may assume, this to be a typical range also for field BSS, implying that the mass accreted is in the range 0.2--0.3\,$M_\odot$. With this perspective any BSS with mass above 1.1\,$M_\odot$ should be considered with caution. In our sample there are potentially three young metal-poor stars: RVS725 ($2.72 \pm 0.34\ M_\odot$; 3.3~$\sigma$ above the 1.6\,$M_\odot$ limit) and RVS726 ($1.80 \pm 0.45\ M_\odot$), already noticed for this reason by \citet{Caffau2024}, and Gaia\_1220 ($1.79 \pm 0.50\ M_\odot$). Most of the other stars younger than 6\,Ga are likely BSS or evolved BSS. Since most paths that lead to the formation of a BSS involve a binary system, it is perhaps not surprising to find several of these in our sample. In fact a high velocity may be acquired in the case of the disruption of a binary system.

\section{Conclusion\label{Sect:Conclusion}}
Using the SOPHIE and UVES spectrographs, we re-observed 134 stars whose radial velocity values, as published in the \gdrthree\ catalogue, are either below $-$500 \kms\ or above $+$500 \kms. We confirm the published values for 104 of these stars, while we refute the measurements of the remaining 30, all of which have \gdrthree\ S/N below~7.

We combined our observations with radial velocity data from the APOGEE, GALAH, GES, LAMOST, and RAVE catalogues to assess the occurrence of erroneous \gdrthree\ radial velocity measurements as a function of S/N and across three absolute radial velocity intervals: [0, 200), [200, 400), and [400, 1000) \kms. The outlier rate reaches up to 83\% in the S/N range [2, 3) and velocity interval [400, 1000) \kms, and decreases rapidly with increasing S/N and/or with decreasing absolute value of the radial velocity.

Finally, the confirmed radial velocities were combined with \gdrthree\ HRV stars having S/N > 10 to construct a clean sample of HRV stars. The majority of these stars follow retrograde orbits. Their positions in the energy-vertical component of the angular momentum diagram coincide with the region where several structures associated with past merging events have been identified: Sequoia \citep{Myeong2019}, Arjuna and I’itoi \citep{Naidu2020}, Antaeus \citep{Oria2022}, ED-2, and ED-3 \citep{Dodd2023}. It is likely that most of these HRV stars were accreted.

Each successive \gaia\ catalogue extends the magnitude limit for which radial velocities are published: $G_\mathrm{RVS} = 12$ in \gdrtwo, 14 in \gdrthree, 15.5--16 in the forthcoming \gdrfour, and a target of 16 or greater for \gdrfive. Despite the acquisition and combination of an increasing number of spectra per stars, the number of very low S/N objects continues to grow steeply with each data release. Identifying and filtering out erroneous values associated with such low S/N measurements remains a key challenge for the future \gaia\ catalogues. For the upcoming \gdrfour\ release, several machine learning techniques have been introduced alongside the more traditional methods used previously, in an effort to improve further the rejection of the spurious measurements.

\section*{Data availability}
Tables A.1 and B.1 are only available in electronic form at the CDS via anonymous ftp to cdsarc.u-strasbg.fr (130.79.128.5) or via http://cdsweb.u-strasbg.fr/cgi-bin/qcat?J/A+A/.

\begin{acknowledgements}
We would like to thank the anonymous referee for their constructive comments, which have helped to improve the clarity and quality of the manuscript. We are grateful to Yveline Lebreton for developing SPInS and for supporting us in its use. This work has made use of data from the European Space Agency (ESA) mission \gaia\ (\url{https://www.cosmos.esa.int/gaia}), processed by the \gaia\ Data Processing and Analysis Consortium (DPAC, \url{https://www.cosmos.esa.int/web/gaia/dpac/consortium}). Funding for the DPAC has been provided by national institutions, in particular the institutions participating in the \gaia\ Multilateral Agreement. This research has also made use of the SIMBAD database \citep{Wenger2000}, operated at CDS, Strasbourg, France. PB acknowledges support   from the ERC advanced grant N. 835087 -- SPIAKID. 
\end{acknowledgements}

\bibliographystyle{aa}
\bibliography{aa56602-25}

\begin{appendix}
\onecolumn
\section{Target coordinates and radial velocities\label{App:Targets}}
Table~A.1 is available online at the CDS (see the Data availability section). It lists the observed targets along with their characteristics. Some were observed on multiple occasions. Each observation is represented by a separate entry, amounting to a total of 169 (for \nbTargets\ stars). For the binary star Gaia\_432, the reported radial velocity corresponds to the mean value of its two components. In Table~A.1, the columns are: (col. 1) \gdrthree\ \texttt{source\_id}, (col. 2) identifier used throughout this paper, (col. 3 and 4) \gdrthree\ right ascension and declination, (col. 5) \gdrthree\ $G$ magnitude, (col. 6) spectrograph employed, (col. 7) Modified Julian Date (MJD) of the start of the exposure, (col. 8) method used to measure the radial velocity, that is cross-correlation -- CC, or template-matching -- TM, (col. 9) radial velocity, (col. 10) radial velocity uncertainty, and (col. 11) status, that is confirmed -- Conf, or spurious -- Spur.

\section{Atmospheric parameters and metallicities for the SOPHIE sample \label{App:AP}}
Table~B.1 is available online at the CDS (see the Data availability section). It contains the atmospheric parameters and metallicities for 50 stars observed with SOPHIE, that had high enough S/N to allow us to derive the metallicity with \mygi\ \citep{Sbordone2014}. For one star, Gaia\_42 (= \gdrthree\ 1992247007192756224) we had two SOPHIE spectra of very different quality (S/N at 500\,nm = 48 and 10 respectively) and we analysed them independently to assess the possible systematic effects due to the S/N. Remarkably the two spectra provide two metallicities that differ only by 0.04\,dex, the lower S/N spectrum providing the higher metallicity. This suggests that we should not expect large systematic effects even in the case of very low S/N spectra. This is in line with the results provided by \citet{Sbordone2014} in their table~5. In Table~B.1, the columns are: (col. 1) identifier used throughout this paper, (col. 2) effective temperature, (col. 3) surface gravity, (col. 4) metallicity, (col. 5) metallicity uncertainty, (col. 6) alpha elements over iron abundance ratio, (col. 7) S/N at 500~nm, and (col. 8) \gdrthree\ \texttt{source\_id}.

\section{Ground-based survey filtering\label{App:GBS}}
To complement our observations, we also made use of five ground-based spectroscopic surveys: APOGEE~DR17 \citep{Abdurrouf2022}, GALAH~DR3 \citep{Buder2021, Zwitter2021}, GES~DR3 \citep{Gilmore2012, Randich2013}, LAMOST~DR7 \citep{Zhao2012, Deng2012, Luo2015} and RAVE~DR6 \citep{Steinmetz2020_1, Steinmetz2020_2}. The following filters were applied to these catalogues:

APOGEE~DR17. Stars were excluded if they did not meet the following criteria: the 4$^\mathrm{th}$ binary digit of \verb|EXTRATARG| set to 0, the 0$^\mathrm{th}$ and 3$^\mathrm{rd}$ binary digits of \verb|STARFLAG| set to 0, the 10$^\mathrm{th}$ and 23$^\mathrm{rd}$ binary digits of \verb|ASPCAPFLAG| set to 0, and \verb|N_COMPONENTS| equal to~1. We also removed duplicate \verb|GAIAEDR3_SOURCE_ID| entries, and the stars with missing values in \verb|VHELIO_AVG|, \verb|TEFF|, \verb|LOGG|, or \verb|FE_H|.

GALAH~DR3. Stars were excluded if they did not meet the following criteria: \verb|snr_c3_iraf|~$\geq$~30, \verb|use_rv_flag|, \verb|flag_sp| and \verb|flag_fe_h| equal to 0. We also removed duplicate \verb|dr3_source_id| entries, and the stars identified as double-line spectroscopic binaries by \citet{Traven2020}, or missing values in \verb|rv_nogr_obst|, \verb|teff|, \verb|logg|, or \verb|fe_h|.

GES~DR3. We removed the stars with duplicate identifiers, or with the \verb|TECH| flag set to 9020, 9030, 9050, 15100, 15110, or 15130, or with the \verb|PECULI| flag set to 2005, 2010, 2020, 2030, 2040, or 2070, or missing values in \verb|VRAD|, \verb|TEFF|, \verb|LOGG|, or \verb|FEH|.

LAMOST~DR7. Stars were excluded if they did not meet the following criteria: \verb|rv_error|~$\geq$~0, \verb|snr|~$\geq$~0, \verb|snrg|~>~25, and \verb|snri|~>~25. We also removed duplicate \gdrthree\ \verb|source_id| entries, and the stars with multiple \gdrthree\ \verb|source_id|, or missing values in \verb|rv|, \verb|teff|, \verb|logg|, or \verb|feh|.

RAVE~DR6. Stars were excluded if they did not meet the following criteria: absolute value of \verb|correction_rv_sparv|~<~10, \verb|correlation_coeff_sparv|~>~10, \verb|hrv_error_sparv|~<~8, \verb|algo_conv_madera|~$\in [0, 2, 3, 4]$, \verb|snr_med_sparv|~$\geq$~20, and \verb|fe_h_chisq_gauguin|~<~1.4. We also removed duplicate \gdrthree\ \verb|source_id| entries, and the stars with multiple \gdrthree\ \verb|source_id|, or with different values of the \verb|flag_1| flag, or with the \verb|flag_1| flag set to \verb|e|, \verb|b|, \verb|p|, \verb|c| or \verb|w|, or missing values in \verb|hrv_sparv|, \verb|teff_cal_madera|, \verb|logg_cal_madera|, or \verb|fe_h_gauguin|. 

\end{appendix}

\end{document}